\documentclass[graybox]{svmult}

\usepackage{mathptmx}       
\usepackage{helvet}         
\usepackage{courier}        
\usepackage{type1cm}        
\usepackage{makeidx}         
\usepackage{graphicx}        
\usepackage{multicol}        
\usepackage[bottom]{footmisc}

\makeindex             

\begin{document}

\title*{Factors in Crowdsourcing for Evaluation of Complex Dialogue Systems}
\authorrunning{Aicher et al.} 

\author{Annalena Aicher, Stefan Hillmann, Isabel Feustel, Thilo Michael, Sebastian M\"{o}ller \and Wolfgang Minker}
\institute{Annalena Aicher \at Ulm University, \email{annalena.aicher@uni-ulm.de}
\and Stefan Hillman \at TU Berlin, 
\email{stefan.hillmann@tu-berlin.de}
\and Isabel Feustel \at Ulm University, 
\email{isabel.feustel@uni-ulm.de}
\and Thilo Michael \at TU Berlin, 
\email{thilo.michael@tu-berlin.de}
\and Sebastian M\"{o}ller
\email{sebastian.moeller@tu-berlin.de}
\and Wolfgang Minker \at Ulm University, \email{wolfgang.minker@uni-ulm.de}
}

%
\maketitle

\abstract{
In the last decade, crowdsourcing has become a popular method for conducting quantitative empirical studies in human-machine interaction. The remote work on a given task in crowdworking settings suits the character of typical speech/language-based interactive systems for instance with regard to argumentative conversations and information retrieval. Thus, crowdworking promises a valuable opportunity to study and evaluate the usability and user experience of real humans in interactions with such interactive systems. In contrast to physical attendance in laboratory studies, crowdsourcing studies offer much more flexible and easier access to large numbers of heterogeneous participants with a specific background, e.g., native speakers or domain expertise. On the other hand, the experimental and environmental conditions as well as the participant's compliance and reliability (at least better monitoring of the latter) are much better controllable in a laboratory.   
This paper seeks to present a (self-)critical examination of crowdsourcing-based studies in the context of complex (spoken) dialogue systems. It describes and discusses observed issues in crowdsourcing studies involving complex tasks and suggests solutions to improve and ensure the quality of the study results. Thereby, our work contributes to a better understanding and what needs to be considered when designing and evaluating studies with crowdworkers for complex dialogue systems.
}



\section{Introduction}
One crucial step in the development of dialogue systems is their evaluation. This evaluation is challenging, as the definition of what constitutes a high-quality dialogue is not always clear and often depends on the specific application, domain, task, and user group~\cite{b2,b15-itu}. Even if a definition is assumed, it is not always clear how it can be measured. For example, if we assume that a high-quality dialogue system is defined by its ability to respond with an appropriate utterance, it is not clear how to measure appropriateness or what appropriateness means for a particular system. Furthermore, one could ask users whether responses were appropriate, but as we will discuss below, user feedback is not always reliable for a variety of reasons. Depending on the ability and purpose of a dialogue system it is unclear how to measure appropriateness or what appropriateness means. Furthermore, the evaluation of dialogue systems is very cost- and time-intensive. This is especially true when the evaluation is carried out through user studies, which compensate users for their participation~\cite{b2}. Therefore, quite a lot of efforts are made, aimed at automating the evaluation, or at least automating certain aspects of the evaluation~\cite{b18-engelbrecht, b19-hillmann, b20-usr, b21-fed}. But still, as automated metrics do not necessarily capture all aspects of the system's quality, a human evaluation is performed, which usually asks about the naturalness and quality of the generated utterances and flow of dialogue~\cite{b5,b16-moeller}. 

One possible way to lower these costs and enable a more flexible evaluation procedure is to perform such human user studies via crowdsourcing. Crowdsourcing engages the help of large numbers of people in tasks, activities, or projects, usually via the internet~\cite{b3}. The application areas of crowdsourcing are very diverse and range from health research areas to the field of market and customer analysis. 

Most studies on crowdsourcing tend to focus on its use of software, technology, online platforms, or its application~\cite{b11-hossain,b12-crequit}. Still, as Bassi et al.~\cite{b8-bassi} point out, there is a need for further exploration to understand how best to use crowdsourcing for research. This is underpinned by numerous references which describe gaps in the research related to crowdsourcing, including a lack of decision aids to assist researchers using crowdsourcing, and best practice guidelines~\cite{b9-buettner,b6-law,b10-sheehan}.

Thus, in the herein presented work, we aim to provide a critical examination and discussion of our experience with crowdworking studies in the context of complex (argumentative) dialogue systems and propose some guidelines for researchers who undertake studies with crowdworkers involving complex tasks in the interaction with dialogue systems.

The remainder of the paper is as follows. Section~\ref{sec:rl} gives an overview of related relevant work and literature. After briefly outlining the essential background information on our crowdsourcing study settings, we discuss the anomalies and problems we encountered in Section~\ref{sec:main}. We introduce recommendations for respective guidelines in Section~\ref{sec:recommendation} and close with a conclusion and a brief discussion of future work in Section~\ref{sec:concl}.

\section{Related Work}\label{sec:rl}
As crowdsourcing has broad application prospects and significant business value, many studies have been conducted in recent years. The existing literature reviews on crowdsourcing range from general overviews (e.g., \cite{b11-hossain,b17-yuen}) to specific research areas like~\cite{b12-crequit}. 
Jiang et al.~\cite{b7} characterize crowdsourcing as suitable for tasks ``that are trivial for humans but difficult for computers, such as classification tasks''. They describe the advantages of crowdsourcing including faster completion speed, lower costs, higher accuracy, and completion of tasks that computers cannot perform. Thereby crowdsourcing engages the help of large numbers of people in tasks, activities, or projects; usually via the internet~\cite{b3}. 

Although a few surveys have attempted to present a more general review of crowdsourcing, they solely introduce definitions of crowdsourcing and/or review typical crowdsourcing systems~\cite{b7}. A highly cited example of a crowdsourcing platform according to ~\cite{b11-hossain} is Amazon’s Mechanical Turk (AMT). An overview of AMT as an academic research platform is provided K.\ B.\ Sheehan~\cite{b10-sheehan} and its strengths (e.g. quick data collection, global respondent pool, relatively low costs, easy handling) and weaknesses (e.g. validity, reliability, ethics) for research are discussed. Fort et al.~\cite{karen} highlight issues raised
by AMT, especially with regard to ethical concerns. Furthermore, Goodman et al.~\cite{b23-goodman} compare AMT participants with other participant samples (community/student) on a set of personality dimensions\footnote{Gosling et al.~\cite{gosling} developed a Ten-Item Personality Inventory (TIPI) which is a 10-item short inventory to assess the big-five dimensions (openness to experience, emotional stability, extraversion, agreeableness, and conscientiousness).} and classic decision-making biases (i.e., framing effects, the conjunction fallacy, and outcome bias). They recommend that researchers should use AMT taking into account certain conditions (integration of screening questions etc.). Manuvinakurike et al.~\cite{b13} present a case study in which 200 remote participants were recruited to participate in a fast-paced image matching game via AMT and the authors discuss encountered technical challenges of the study.

Beyond technical challenges also potential problems and issues are discussed among some crowdsourcing-related publications. For instance, Thebault-Spieker et al.~\cite{b14} discuss how mobile crowdsourcing tasks, situated in the physical world (e.g., checking street signs, running household errands), are influenced by geographical implications (suburbs, etc.). Nevo et al.~\cite{b1-nevo} state that most crowdsourcing experts found that ``stand-alone tasks'' and tasks with a ``clear definition'' are best suited for crowdsourcing. Furthermore, they claim that crowdsourcing projects require specific domain knowledge and a ``well-developed problem statement''.
Nevertheless, Noel-Storr et al.~\cite{b3} stress that there is evidence that a ``crowd'' of non-specialists is capable of tasks like identifying quantitative studies by topic-based citation screening (i.e., assessing titles and abstracts). Thus, there seems to be a dependency on the expertise, skills, and knowledge of crowdworkers and the complexity and difficulty of the task. This is underpinned by our findings as well and further discussed in Section~\ref{sec:main}. Furthermore Shumeli et al.~\cite{shumeli} draw attention to the lack of ethical considerations related to the various tasks performed by workers, including labeling, evaluation, and production. 

Another crucial aspect we also have strongly experienced in our crowdsourcing studies is described by Palogiannidi~\cite{palogiannidi2013using}. She claims that crowdsourcing's main advantage---the great diversity of people groups that participate --- is also its main disadvantage. They refer to the supposed ``tendency of humans to cheat, whenever it is possible, especially if anonymity is preserved'' (ibid. 30). Furthermore cheating was detected in a spoken dialogue system's related crowdsourcing task (assessment of text-to-speech synthesis) by Buchholz and Latorre~\cite{b28-buchholz}.

To counteract these disadvantages as far as possible and design tasks that will yield a higher quality of the workers’ responses Palogiannidi~\cite{palogiannidi2013using} stresses the importance to quantify senses such as freedom, politeness, specificity, etc during the design process of crowdsourcing tasks. On the one hand, the selection of the crowdworkers can be adjusted, especially with regard to their expertise and suitability for a task. Liao et al.~\cite{b27-liao} proposed an approach that aims to recommend a group of suitable workers through worker personality analysis and community classification. On the other hand, it makes sense to define guidelines and best practice recommendations to conduct crowdsourcing studies, which is also our main objective in this paper. For instance, Sabou et al.~\cite{b22-sabou} discuss crowdsourcing methods for corpus acquisition and propose a set of best practice guidelines based on their experiences. Furthermore, Ramírez et al.~\cite{b29-ramirez} examined the current state of reporting of crowdsourcing experiments on the basis of 171 experiments and propose a checklist for crowdsourcing experiments. 

With regard to (spoken) dialogue systems for instance Li et al.~\cite{lietal2021} introduce LEGOEval, an open-source Python-based toolkit that allows researchers to easily develop human evaluation tasks for dialogue systems on AMT in a LEGO plug-and-play fashion. In~\cite{juricec2011} Jurcicek et al. discuss a framework for evaluating dialogue systems using Amazon Mechanical Turk for recruiting a large group of users and conclude that crowdsourcing provides an effective method of rapidly testing spoken dialogue systems at a modest cost. Furthermore, Huynh et al.~\cite{huynhetal2022} stress the importance of the clarity of instructions, examples, fair payment, and low quality to be considered when creating Human Intelligence Tasks~\footnote{AMT terminology for tasks like image labeling,
semantic labeling and audio transcription.} so that the data gathered is of the highest quality possible. But the existing literature concerning very advanced and complex tasks, i.e. real-time interactions over a lot of turns with a complex (spoken) dialogue system is very scarce.

According to Buettner~\cite{b9-buettner} the majority of the crowdsourcing papers that discuss complex and creative tasks are related to idea generation, competition, and evaluation by the crowd. Thus in this paper, we aim to close this gap and report our positive and negative experiences with crowdsourcing-based user studies considering complex dialogue systems.
To the best of our knowledge, so far there does not exist a respective discussion of crowdsourcing studies for complex (argumentative) dialogue systems nor a suggestion for guidelines to perform such studies. 

\section{Studies}\label{sec:main}
In this section, we will describe the study settings and evaluate our findings of three different user studies conducted via crowdsourcing. Each study was performed with a different type of dialogue system, indicating that the tasks and their respective complexity were strongly dependent on the respective dialogue system.

\subsection{Study 1: Comparison of Modalities in Argumentative Dialogue Systems}
The aim of this study was to analyze the impact of different input/output (I/O) modalities (I: drop-down menu / O: text vs. I: speech / O: speech) on the evaluation of an argumentative dialogue system (ADS) and user interest.

\subsubsection{Study Settings}
Our user study~\cite{lrec-eval} was conducted online via the crowdsourcing platform \textit{Crowdee}\footnote{\url{https://www.crowdee.com/}} with participants from the UK, US, and Australia in the period 12-29\textsuperscript{th} November 2021. All 292 crowdworkers were non-experts without a topic-specific background. The crowdworkers could access the study via their \textit{Crowdee} job board, which showed the title ``Use our BEA dialogue system and provide feedback!'', the language option English and the estimated task duration. After the job was selected the task description: ``to listen to enough arguments to build a well-founded opinion~\footnote{BEA is a cooperative argumentative dialogue system that has neither persuasion nor its own (ethically critical) agenda. It merely presents the users with arguments on a certain topic on their demand and gives them the opportunity to express their opinion.} on the topic ``Marriage is an outdated institution''\footnote{This sample debate serves as knowledge base for the arguments and is taken from the \textit{Debatabase} (\url{https://idebate.org/debatabase}, last accessed 19\textsuperscript{th} November 2021).} (at least 10 arguments)'' was displayed. If the participants confirmed their willingness to proceed and accepted the data protection regulation guidelines they were redirected to the study introduction page. 
There, a short text and demo video (menu: 1:32min, speech: 2:02min) explained how to interact with the assigned version of the dialogue system (either menu-/GUI-based or speech-based input modality). Particular attention was paid to the speech interaction system, where the users had to express their requests in their own words, whereas, in the menu system, they had to click on a response option. Therefore, the speech system represented a significantly more demanding and complex setting as the user had to understand the arguments, how to interact with the system, and formulate the respective requests. Thus, in the speech system participants get an instruction how about the actions they can perform and how to express them. Furthermore, users of the speech system could ask for assistance during the interaction if they were unsure about what to do next. In response, the system displayed all possible interaction options on the graphical user interface of the dialogue system.
One of the main objectives of this study was to estimate and analyze changes in the user's content-related interest. Therefore, the users were asked to rate their interest in the currently presented aspect/argument as described in detail in Aicher et al.~\cite{lrec-eval}. In case the users lost interest in participating/continuing with the task, it was repeatedly stressed that the end of the interaction could be chosen freely as soon as at least ten arguments were heard. The minimum amount of arguments was chosen to ensure that enough data could be collected. 
The first 139 participants interacted with our ADS via drop-down menu input, and the other 153 via speech. The latter user group was strongly advised to use headsets to reduce background noise. Each participant completed the study online without any supervision. Before and after the conversation with the system the participants had to rate statements (meta questions regarding the interaction, e.g. user's perception and impression of the system etc.) on a five-point Likert scale. A part of these questions was taken from a questionnaire according to ITU-T Recommendation P.851\footnote{Such questionnaires can be used to evaluate the quality of speech-based services.}~\cite{b30-itut}. Furthermore, these questionnaires contained three control questions, to identify participants who did not fill out the questionnaire seriously or carefully, e.g., by randomly selecting options on the Likert scale~\cite{b23-goodman}. In particular, Aker et al.~\cite{aker2012assessing} stress that real tasks require high precision objective questions such as the maths questions in combination with the free text design can be used to filter out non-compliant or ``unprecise workers''.
Retrospectively, it became clear that establishing some demographic facts of the participants (gender, age,..) would have been helpful for completeness and for the interpretation of the results, which should therefore be included in further studies.  

The evaluation of these control questions and considering additional feedback revealed that the data of 90 participants showed anomalies. Their data were excluded according to previously defined exclusion criteria: Contradictory answers in control questions in the questionnaire, taking less than 30 seconds to read through the introduction and watch the introduction videos, taking less than 120 seconds to answer the 40+15 questions in the final questionnaire and feedback indicating that problems occurred during the interaction or participants reported that they did not know what to do. This led to a total number of data records of 202 participants (menu: 104, speech: 98) which were included in the analysis.

\subsubsection{Evaluation}
On average the participants interacted with the ADS for 31:45 minutes (menu: 27:57, speech: 35:34). This significant difference can be explained by the fact that the spoken interaction (speaking and hearing) inherently takes longer than clicking on an option in a drop-down menu and reading the response. This also displays one of the disadvantages of the menu system as users do not have to attentively read the presented argument but can just move on by clicking the next move, whereas in the spoken interaction, the user is forced to listen to the ADS.

But even though the average time the users of the menu system interacted with the ADS is lower, the number of provided arguments is significantly higher in the menu system. 9.6\%/17.3\% of the menu/speech system participants quit the conversation after hearing the minimum number of arguments (in total: 13.4\%). Most of the participants heard between 20-30 arguments of 72 available arguments, which indicates that even though the given task was quite subjective, the crowdworkers interacted with the system longer than they had to\footnote{The users could take as much time as they wanted for each response, as well as for the overall interaction.}. These findings underpin the ones of Nevo et al.~\cite{b1-nevo}, who described a problem with the strict duration of specific steps, which is unlikely to occur if the participants are not bound to any stepwise time limits during the interaction. 

Regarding all aspects of the evaluation questionnaire, the menu system outperformed the speech system significantly. This was particularly observed in aspects that rated if the system provided the expected information or errors in the interaction which occurred. 
To a certain extent, this points to a lack of processing of the user utterances and can either be explained by errors in the automated speech recognition (ASR) or the natural language understanding (NLU) module of the ADS. By checking the dialogue logs of the interactions with users in the speech system, we found that about 15\% of all spoken utterances were recognized erroneously by the ASR. Even though the NLU module matched about a third of these erroneously recognized utterances to the correct (i.e., intended) action, the impact is still significant. The errors of the ASR might partly be explained by the participants not wearing headsets during the interaction which led to additional noise.
In order to avoid or mitigate problems due to insufficient speech or audio quality one could either switch to typed input or in order to preserve the speech modality and ensure suitable environmental conditions by conducting the study in the lab on site. It is also possible to check the audio quality on the user's device, but this is technically complex (i.e. error-prone), detects problems \textit{after} the user has started the task and prolongs the overall task.
 
Still, there are some observations that strongly suggest that this outcome was also influenced by the unsupervised introduction to the task\footnote{Unfortunately the introduction could be skipped rather easily}. We noticed inconsistencies in the user behavior, as users repeated their request multiple times, without reacting to the system's answer to choose another action. This indicates that in contrast to the menu system, where the users were always only displayed the available option, the speech users had to figure out what actions they can perform and how to suitably formulate it. Even though the system's design incorporated a ``Help'' button, as well as the ``available'' options action, only 15\% of the speech users used them. This can be explained by the fact that only 35\% of the users spend enough time on the introduction website to read through the explanation and watch the according video properly.

Unfortunately, we did not include any test questions before the interaction to check if the speech participants thoroughly read/viewed the interaction instructions. Thus the speech system displayed a far more complex task whereas the menu system just displayed the actions the user had to choose from. Even though the latter system presented less flexibility and was perceived as static and not natural, the speech system was perceived as too complicated and unreliable.
This is underpinned by feedback that was received from the crowdsourcing platform~\textit{Crowdee} after the study, e.g.,  stating that ``It was not possible to do what I wanted to do. I repeated myself many times''/``I was stuck in the argument and could not get back''/``The idea is cool, still, sometimes things got messy when I could not proceed to the argument I wanted to go to.''. Therefore, it might useful to conduct a double-staged study, which ensures that all participants passing the first stage have well-understood how the interaction with the ADS works. Another option would be to conduct on-site, where the participants can be supervised and the environmental conditions are controlled.

A further indicator of the higher complexity of the speech system can be seen in the number of study abortions, as it needed seven times more participants to achieve the targeted number of participants.

In general, the results clearly show that the input/output modalities and respective difficulties/problems decrease the rating of the general impression of the system, even in aspects that have no relation to the former. For example, the incremental approach to present arguments, the sufficiency of different options, or the conclusiveness of arguments, which depend on the content but not on the modality, are rated in the speech system significantly worse than in the menu system. Therefore, it is crucial to solve the identified issues and take precautions, e.g., in the design of crowdsourcing studies to avoid them.

\subsection{Study 2:  Evaluation of a Pipeline for Argumentative Dialogues}
In order to introduce arguments into dialogue systems in a meaningful way, a suitable data structure for the dialogue flow is required. In our previous works~\cite{b26-rach} a tree structure that allows access to arguments in dialogue form was developed. The tree represents both pro and con arguments to a statement and their relation to each other. Until now, these tree structures were created manually requiring a lot of effort. Therefore, we~\cite{b26-rach} developed a pipeline to feed arguments from an argument mining search engine into a tree structure suitable for argumentative dialogues. These structures were then used in a agent-agent dialogue system, where each agent had either a pro or con position on the discussed topic. In order to test the pipeline, several studies were conducted, which were mainly intended to check the (logical) coherence of the arguments and the overall dialogue. 

\subsubsection{Study Settings}
In order to measure coherence, three categories were used: Comprehension, Reference, and Polarity. Each category was converted into a yes/no question that directly assessed the utterance properties that were affected by the retrieval pipeline \cite{b26-rach}.
The resulting questions are as follows:
\begin{itemize}
    \item Comprehensible: Do you understand what the speaker wants to say?
    \item Reference: Does the utterance address its reference?
    \item Polarity: Does the utterance contradict the speaker's position?
\end{itemize}
The web interface allowed participants to rate utterances of the generated dialogues with regard to the mentioned categories.
Before, an introduction to the study was provided containing a sample dialogue with an exemplary rating and respective explanations. Each participant had to evaluate three dialogues. Afterwards, a questionnaire on the comprehensibility of the categories had to be answered, as well as a text field for further comments on the study.
The dialogues for the study were created from argument structures for seven different topics with two different configurations of the pipeline and one annotated structure. A total of 30 dialogues with equal amounts of utterances were created which were divided into 10 groups. Each group consisted of seven participants, such that a total of 70 participants was needed. The study was conducted via the crowdsourcing platform \textit{Clickworker}\footnote{\url{https://www.clickworker.com}}.

\subsubsection{Evaluation}
A total of 10122 ratings for all three categories were collected with the study. Since the ratings of the participants diverged strongly, the three most agreeing users of each group were chosen for the evaluation. Results were found to be highly subjective - with regard to the direct comparison to each other and the impact of the pipeline configuration.

Considering the questionnaire responses in terms of the comprehensibility of each category, minor problems could be perceived. Unfortunately, 29 participants rated at least one category as incomprehensible. The majority of the ratings are neutral or positive, nevertheless, comprehension problems seemed to have been present, especially with regard to the polarity question. This was also confirmed in the test stage of the study. A few test participants were asked to test the study extensively. Although they were partially experts with knowledge of the argumentation domain, they found it difficult to evaluate the study correctly in all points. This strongly indicates the high complexity of the task even for experts. Furthermore, determining the coherence of arguments is a challenging and difficult task. 

We assume that the participants of the ``real'' study without expert knowledge may have needed a supervised introduction to fully understand the categories and their usage. Nevertheless, it is not clear whether this would have led to a better result than the crowdsourcing study. Since this study primarily consisted of a questionnaire setting rather than a real conversational interaction, the use of crowdsourcing seems still adequate.

\subsection{Study 3:  Data Acquisition for Indoor-Navigation Dialogue System}
Further insights into crowdsourcing-related issues are provided by our data collection study~\cite{b25-miehle-feustel}. The purpose of this study was to collect route descriptions using different communication styles for an indoor navigation dialogue system. Therefore, we conducted a crowdsourcing study, which showed a video with a route, which the study participants were then asked to describe.

\subsubsection{Study Settings}
We chose three routes in a public university building and filmed videos while walking along the former. All of the routes contained different points of interest and route elements like the cafeteria, an elevator, or stairs. The start and end points of the routes were linked such that all routes could be taken one after the other. In other words, route one ends where route two starts, and the end point of route two is the starting point of route three.
In order to perform the study, a web application containing a few demographic questions (age, gender, and country of residence) and a video description tool were implemented. The tool consists of two parts: the video and the description box. Within the description box, the users had to enter their route descriptions while watching the video. While typing, the video stopped automatically and when pressing enter to save the input, which was then added to a list, the video continued. All saved items could still be edited or deleted. To prevent misuse, a minimum amount of items was required for the video to proceed in combination with a timer which secured that the video had been watched completely.
The study was conducted via the crowdsourcing platform \textit{Clickworker} allowing for English native speakers. We aimed to have 100 participants, 50 from the US and 50 from the UK (each with 25 females/males).

Prior to starting the study, three users from Germany checked and tested the system with regard to smooth operation, potential errors, comprehensibility of the task, etc.

\subsubsection{Evaluation}
Overall we had 97 participants, 53 from the UK (28 female, 25 male) and 44 participants from the US (18 male, 25 female, 1 prefer not to tell).
Unfortunately, some of the participants misused our study tool and did not fill in the correct data. Moreover, some participants did not finish their descriptions. The distribution of participants with respect to the quality of their data is described in the following:
\begin{itemize}
    \item 74 participants entered correct data that could be used after minimal changes (e.g., merging some items due to unnecessarily created items).
    \item  8 participants provided correct information but did not finish the description. We assume that the participants did not see a need to continue the study after the timer expired. Nevertheless, this data could still be used as it is basically correct although incomplete.
    \item 15 participants entered invalid data (e.g., wrong route descriptions like left instead of right) or data that required a lot of post-processing and correction.
    \item 3 participants received the compensation for crowdworkers without even entering data. We assume that the code for the payment was passed by other users.
\end{itemize}
When evaluating the study data, we encountered two main challenges. The first is that there were many attempts to abuse the study setup, despite some precautions (e.g., time-based buttons to proceed according to the video length). Nevertheless, 18\% of the study data had to be excluded (invalid/no data). In general, it was very time-consuming to filter out incorrect data records, since the entire route had to be traced in order to find an incorrect description. Second, all statements were written in chat style and not necessarily in text for spoken language, therefore the statements had to be partially expanded to fit our purpose. The adjustments of the statements in order to be able to use them for our dialog system were therefore indispensable. Nevertheless, as we aimed for a study with English native speakers a crowdsourcing study was suitable for this task. Moreover, it was crucial to access a large number of heterogeneous participants (only possible via crowdsourcing) in order to capture different communication styles. Thus, we already expected additional effort for data post-processing, since the extraction of the communication styles had to be done manually. 

To conclude, we found that not only to summarize the different study results presented here and to derive respective general recommendations, but also to generally guarantee comparability and reproducibility of the results of crowdsourcing studies, a certain standardization, e.g. using checklists~\cite{b29-ramirez}, would be very helpful.

\section{Recommendations}\label{sec:recommendation}
Based upon our experiences and analysis of the study results described in Section~\ref{sec:main} we propose the following recommendations for researchers using crowdsourcing to evaluate dialogue systems tailored particularly for complex tasks.

\begin{itemize}
    \item Incorporate measures to ensure that the necessary instructions such as the introduction text and instructional videos can not be skipped, e.g., respective timers, test questions that can only be answered with careful consideration, etc. .
    \item Provide various options for assistance, which are easy to notice. Offer proactive assistance and more explanations, if the interaction is paused for a certain period of time or the user obviously does not know what to do (performing actions that are not possible, repeating an action without taking the system's reaction into account).
    \item Double-staged study setting: Incorporating simulated test requests/actions in a first step, which emulate the possible actions the user can perform in the real interaction. Thus, it can be checked whether the participant understood the assignment and is only allowed to proceed to the real interaction if executed correctly. 
    \item Record and consider demographic data and differences, especially with regard to argumentative dialogue systems and controversial topics.
    \item Include screening questions that gauge attention and language comprehension and can be used to filter out random ticking~\cite{b23-goodman} and not conscientious participants. 
    \item Usage of checklists~\cite{b29-ramirez} when reporting crowdstudies to ensure reproducibility and comparability of results (post self-control).
    \item Always use unique, one-time redeemable payment codes for participants compensation. 
\end{itemize}

\section{Conclusion and Future Work}\label{sec:concl}
In this work, we have described and analyzed three crowdsourcing studies of three different dialogue systems and have identified issues related to crowdsourcing. Based upon this we propose recommendations for the design of crowdsourcing studie with dialogue systems, in particular, if the latter involve more complex task. Furthermore, if the complexity exceeds a certain level, we recommend to perform user studies under supervision of experimenter on site who can assist accordingly. 

In future work, we will incorporate the recommendations in Section~\ref{sec:recommendation} in our crowdsourcing study setup and analyze whether this solves or reduces the described issues, respectively. Furthermore, we plan a rerun of Study 1 (see \textit{A.} in Section~\ref{sec:main}) in the lab to further investigate the influence of the study environment (lab vs. crowdsourcing) on the study evaluation results. Especially, we are interested in how the interaction, the perception of the systems and respective user ratings change under direct supervision. 

\section*{Acknowledgment}
This work has been funded by the DFG within the project ``BEA - Building Engaging Argumentation'', Grant no. 313723125, as part of the Priority Program ``Robust Argumentation Machines (RATIO)'' (SPP-1999).





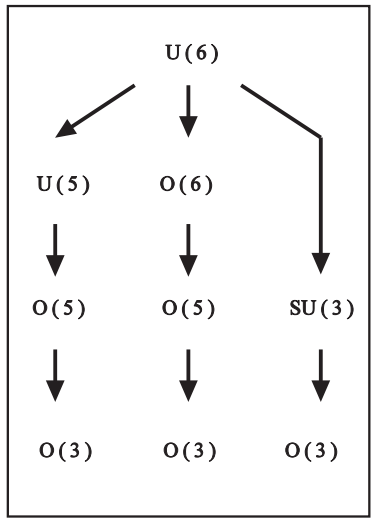
\end{document}